\documentclass[conference]{IEEEtran}
\IEEEoverridecommandlockouts
\newcommand{\h}[0]{\hspace{2ex}}
\newcommand{\spacetune}[1]{#1}

\newcommand{\code}[1]{{\texttt{#1}}}
\newcommand{\quotett}[1]{``#1''}
\usepackage{graphicx}
\usepackage{comment}
\usepackage{balance}
\usepackage{lipsum}
\usepackage{phoenician}
\usepackage{polyglossia}
\setmainlanguage{english}
\newfontfamily\runicfont{FreeMono.otf}
\usepackage{xcolor}
\definecolor{linkc}{rgb}{0.1,0.1,.8}
\definecolor{darkgreen}{rgb}{0,0.5,0}
\definecolor{midblue}{rgb}{0,0,0.7}
\usepackage{tikz}
\usetikzlibrary{shapes.geometric, arrows, positioning, matrix,arrows.meta,calc}
\definecolor{linkc}{rgb}{0.1,0.1,.8}
\definecolor{darkgreen}{rgb}{0,0.5,0}
\definecolor{midblue}{rgb}{0,0,0.7}
\definecolor{darkblue}{rgb}{0,0,0.5}
\definecolor{darkred}{rgb}{0.5,0,0}
\definecolor{titleblue}{rgb}{0.08,0,0.5}
\usepackage{listings}
\lstdefinelanguage{futhark}
{
  morekeywords={
    alloc,
    scan,
    map,
    map2,
    map3,
    map4,
    concat,
    def,
    do,
    else,
    f32,
    u32,
    bool,
    for,
    fun,
    if,
    in,
    include,
    let,
    loop,
    struct,
    then,
    type,
    val,
    while,
    with,
    mapnest,
    module,
    where,
    sort,
    scratch,
    multired,
    reverse,
    zip3,
    unzip3,
    copy,
    gather,
    withacc,
    sum
  },
  sensitive=true, 
  morecomment=[l]{--}, 
  morecomment=[s]{\{-}{-\}}, 
  morestring=[b]", 
  literate={\\}{\fn}{1} {->}{$\rightarrow$}{1} {<-}{$\leftarrow$}{1} {|>}{$\pipe$}{1} {<|}{$\revpipe$}{2}, 
}

\usepackage{xcolor}
\definecolor{eclipseBlue}{RGB}{42,0.0,255}
\definecolor{eclipseGreen}{RGB}{63,127,95}
\definecolor{eclipsePurple}{RGB}{127,0,85}

\newcommand{\pipe}{\ensuremath{\triangleright}}
\newcommand{\revpipe}{\ensuremath{\triangleleft}}

\usepackage{url}
\usepackage[bookmarks=false,colorlinks=true,citecolor=darkgreen,linkcolor=linkc,filecolor=linkc,urlcolor=linkc]{hyperref}

\usepackage{amsmath}
\usepackage{algorithmic}
\usepackage[caption=false,font=footnotesize]{subfig}
\usepackage{cleveref}

\usepackage[inline]{enumitem}
\usepackage{cite}      
\usepackage{array}     
\newcommand{\cgbn}{\textbf{\sc cgbn}}
\newcommand{\cuda}{\textbf{\sc cuda}}
\newcommand{\soac}{\textbf{\sc soac}}
\newcommand{\cpp}{\textbf{\texttt C++}}

\begin{document}

\author{
\IEEEauthorblockN{Cosmin E. Oancea}
\IEEEauthorblockA{
DIKU, University of Copenhagen \\ Copenhagen 2100, Denmark\\
\texttt{cosmin.oancea@di.ku.dk}}
\and
\IEEEauthorblockN{Stephen M. Watt}
\IEEEauthorblockA{Cheriton School of Computer Science, U. Waterloo \\
Waterloo, Canada\\
\texttt{smwatt@uwaterloo.ca}}
}

\title{\mbox{High-Level Big Integer Arithmetic} \mbox{in Futhark for GPUs}\\
\thanks{Research reported here was partially supported by the Novo Nordisk Foundation award NNF24OC0090447 and the Natural Sciences and Engineering Research Council of Canada.}
}
\date{}
\maketitle
\begin{abstract}
We report on GPU implementations of block-level
addition, subtraction, multiplication and division for
midsize integers, with operands of $2^{15}$ to $2^{19}$ bits
using the high-level functional language Futhark.
Comparing with hand-written \cpp{}/\cuda{}
versions and \cgbn{}, we identify which functional
constructs compile well, where memory placement and
sequentialization are effective, and what compiler support
is needed.  
The results show that high-level code can express
the algorithms compactly while approaching competitive
performance after certain compiler improvements.
In particular, we find that automated placement of arrays in GPU register memory is critical for performance.
\end{abstract}

\section{Introduction}

\spacetune{\enlargethispage{\baselineskip}}

We examine the use of a high-level functional language,
Futhark~\cite{futhark-ppopp}, to produce efficient general purpose GPU code for big integer arithmetic. In earlier work
\cite{OanceaWatt2025,OanceaWatt2026a}, we gave algorithms
for addition, subtraction, multiplication and division
suitable for integers with $2^{15}$ to $2^{19}$ bits. These
were implemented in \cpp{} using \cuda{}, but our longer-term
objective has been to obtain high-level functional implementations
with competitive performance.

Our motivation is that high-level algorithms can be rapidly
deployed and modified, can expose optimization opportunities
at a higher semantic level, and can clarify the cost of
abstraction. In particular, we ask:
\begin{itemize}
\item Can higher order functions and automated function
fusion give efficient arithmetic on GPUs?
\item Can automated memory allocation and object placement
be made sufficiently efficient?
\item What primitives are needed for an elegant and efficient
solution to a classical problem --- big integer arithmetic?
\item Can such an implementation provide a competitive
library for GPU applications?
\end{itemize}
More generally, we would like to know how much
machine-specific detail can be left to a high-level language compiler without
losing the essential performance characteristics of the
underlying algorithms. This paper provides first answers to
these questions.

GPU arithmetic has
been studied from low-level multi-precision integer libraries
to exact polynomial arithmetic for computer algebra. 
We highlight here some work relevant 
to the design choices in this paper. 
The most
directly comparable integer package is NVIDIA's Cooperative
Groups Big Numbers library, \cgbn{}~\cite{CGBN}, which maps
each arithmetic instance to a small cooperative group,
typically no larger than a ``warp'' (see
Section~\ref{sec:machine-model}). It exploits fast warp-level
communication and gives excellent performance for fixed
precisions in its intended range, but is less well matched to
the larger regime considered here. Other GPU integer work has
explored hardware-adder-style carry propagation, tiled
quadratic multiplication, and FFT-based multiplication using
real transforms or GPU libraries
\cite{EmmartWeems2010,DieguezEtAl2022,Bantikyan2014}.
Related multiple-precision floating-point libraries include
CUMP and CAMPARY~\cite{NakayamaTakahashi2011,CAMPARY2016};
they address a different numerical problem, but illustrate
the long-standing interest in GPU arithmetic beyond machine
precision.

A related line of computer-algebra work studies dense
univariate polynomial arithmetic over finite fields. 
Earlier work used GPU FFTs over finite fields for polynomial
multiplication~\cite{MorenoMazaPan2010} 
and later showed that, for important degree ranges, plain
multiplication, division and GCD can outperform FFT methods
once GPU overheads are included~\cite{HaqueMorenoMaza2012}.
CUMODP developed this direction into a \cuda{} library for
exact dense polynomial arithmetic over finite fields,
including subresultants, multipoint
evaluation and interpolation~\cite{CUMODP2014}. 
Later
work studied GPU transforms over large prime fields, including generalized Fermat prime
characteristics~\cite{ChenEtAl2017PrimeFFTGPU}.
These computations share with integer arithmetic convolution-like
products, finite-field transforms, and tradeoffs among
asymptotic work, memory traffic, synchronization and
kernel-launch overhead. Integers add the complication that
carries and borrows must be propagated.

\spacetune{\enlargethispage{\baselineskip}}

The present paper asks how well these algorithms can be
expressed in Futhark, a high-level purely functional array
language, and what compiler support is
needed for such code to approach hand-written \cuda{}
performance.
The remainder is organized as follows: Section~\ref{sec:machine-model}
describes the GPU model; Section~\ref{sec:futhark}
reviews Futhark; Section~\ref{sec:compiler-improvements} discusses compiler improvements related to automatic placement in register memory; 
Section~\ref{sec:futhark-impls} presents the algorithms and their Futhark implementations; Section~\ref{sec:evaluation} evaluates performance and Section~\ref{sec:conclusions} concludes.

\section{Machine Model}
\label{sec:machine-model}
Our algorithms are expressed in a high-level language
compiled to a vendor-neutral GPU model. We use the
OpenCL/SYCL terminology: a computation is launched as a set
of lightweight \quotett{work-items}, organized in
\quotett{work-groups}. In \cuda{} terminology, these
correspond to \quotett{threads} and
\quotett{thread blocks}. Work-groups are scheduled on GPU
compute units or analogous execution resources. Work-items
within a work-group may synchronize and use local memory;
different work-groups have no portable synchronization
within a kernel and communicate through global memory and
kernel boundaries.

Global memory is large but slow.
Local memory,
called \quotett{shared memory} in \cuda{} and
\quotett{local memory} in OpenCL/SYCL, is scoped
to one work-group and is much faster. Private storage,
usually registers, is fastest and belongs to one work-item. 
Efficient programs reuse data in local and/or private storage before writing back. Limited fast storage determines occupancy and problem size. 

In our work, a multi-precision integer is an array of $M$
fixed-width digits in base $B=2^b$, stored in little-endian
order, where $b$ is the digit width. We target midsize
integers for which one arithmetic instance fits in one
work-group: beyond single-sub-group methods, but small
enough for operands and key intermediates to reside in fast
storage. A logical digit need not correspond to one
work-item: each work-item may process $Q$ logical elements,
so $T$ work-items cover $QT$ digit positions. 
This sequentialization reduces synchronization, improves register
reuse, and permits larger integers, at the cost of register
pressure.

Thus we rely only on work-group synchronization, local
memory, sub-group execution, coalesced global memory, and
scarce fast storage. This is weaker than a model exposing
vendor-specific primitives, but better matches
the portability goals of a high-level language
implementation.

\section{Futhark}
\label{sec:futhark}

\begin{figure}
\begin{lstlisting}[numbers=none,language=futhark,mathescape=true,xleftmargin=\parindent]
def iota (n: i64) : [n]i64 = ...-- [0,1,$\ldots$,n-1]
def replicate 't (n:i64) (v:t) : [n]t =...-- [v,$\ldots$,v]
def zip  [n]'$\alpha$'$\beta$ (a: [n]$\alpha$) (b: [n]$\beta$) : [n]($\alpha,\beta$)= $\ldots$
def unzip[n]'$\alpha$'$\beta$ (aos: [n]($\alpha,\beta$)) : ([n]$\alpha$, [n]$\beta$)= $\ldots$
def map [n] '$\alpha$ '$\beta$ (f: $\alpha \rightarrow \beta$) (a: [n]$\alpha$) : [n]$\beta$
           --  [ f a[0], $\ldots$, f a[n-1] ]
def map2[n]'$\alpha$'$\beta$'$\gamma$(f:$\alpha \rightarrow \beta\rightarrow\gamma$)(a:[n]$\alpha$)(b:[n]$\beta$): [n]$\gamma$
      -- [ f a[0] b[0], $\ldots$, f a[n-1] b[n-1] ]
def scan[n]'$\alpha$ ($\odot$:$\alpha \rightarrow \alpha \rightarrow \alpha$)($ne_\odot$:$\alpha$)(a:[n]$\alpha$) : [n]$\alpha$
    -- [ a[0], a[0]$\odot$a[1], $\ldots$, a[0]$\odot~\ldots~\odot$a[n-1] ]

loop $\overline{x}$ = $\overline{x^{init}}$ for i < $e^n$ do $e^{body}$(i,$\overline{x}$)  $\equiv$  f 0 $\overline{x^{init}}$
 $\textbf{where}$ f i $\overline{x}$ = if i==$e^n~$then$~\overline{x}~$else f (i+1) $e^{body}$($\overline{x}$,i)

def write[n]'$\alpha$(x:*acc([n]$\alpha$))(i:i64)(v:$\alpha$):*acc([n]$\alpha$)
def reduce_by_index_stream [k]'$\alpha$'$\beta$ (dest: *[k]$\alpha$)
 ($\odot$:$\alpha\rightarrow\alpha\rightarrow\alpha$)(ne$_\odot$:$\alpha$) (f:*acc([k]$\alpha$)$\rightarrow\beta\rightarrow$acc([k]$\alpha$))
 (bs: []$\beta$) : *[k]$\alpha$ = $\ldots$
def scatter_stream [k]'$\alpha$'$\beta$ (dest: *[k]$\alpha$) 
  (f: *acc([k]$\alpha$) -> $\beta$ -> acc([k]$\alpha$)) (b:[]$\beta$): *[k]$\alpha$=$\ldots$

-- Example of using scatter_stream, write and loop:
def shift[m][q] (n:i64)(xs:[m][q]uint): *[m*q]uint=
  let f (A: *acc([m*q]uint)) tid : acc([m*q]uint) =
    loop A for i < q do
      let off = q * tid + i + n
      let (index,value) = if off >= 0 && off < m*q
         then (off, xs[tid,i]) else (m*q-off+n-1, 0)
      in  write A index value
  in scatter_stream (replicate (m*q) 0) f (iota m)
\end{lstlisting}\vspace{-1ex}
\caption{Futhark constructs \& demonstration on shifting a 2D array}
\label{fig:futhark-soacs}
\vspace{-\baselineskip}
\end{figure}

{\em Futhark}
is a purely functional array language described elsewhere~\cite{futhark-ppopp}. 
It is named after Elder Futhark, the oldest known runic script, believed to have originated in the region of present-day Denmark. 
The name comes from ``{\runicfont ᚠᚢᚦᚨᚱᚲ}'', the first six runes of its alphabet,
much as ``alphabet'' is from the first two Greek letters.

Here we give a few minimal points on syntax to make programs legible.
Function application uses the notation popularized by ML, so \texttt{f a b} is \texttt{f} applied to \texttt{a}, giving a function then applied to \texttt{b}.   This does not preclude giving a single tuple-valued argument, \textit{e.g.} \texttt{f(a,b)}.
The pipe operator, written $\triangleleft$ or \texttt{<|}, gives the value on the left as an argument to the right, so \texttt{~a <| f~} is \texttt{~f(a)}.

\spacetune{\enlargethispage{\baselineskip}}
Data-parallel array languages---APL~\cite{hsu2019data,DBLP:conf/pldi/HsuS23}, Accelerate~\cite{accelerate-first,accelerate-nested,accelerate-scans}, DaCe~\cite{dace-stateful,dace-climate,dace-quantum-transport}, Futhark, JAX~\cite{jax1,jax2}, Lift~\cite{Lift,steuwer2022rise}, SAC~\cite{GrelckScholz05,GrelTrojIFL04,GrelSchoPARCO06}---express computations by nesting and composing second-order array combinators, such as \code{map}, \code{reduce}, \code{scan} (prefix sum), \code{scatter} (un-structured write), and sequential loops as well. In short, they aim to support parallel-correctness guarantees and an elegant expression that is close to the algorithm, while still offering good performance. We refer the reader to~\cite{cfal} for a comparative study of some of these languages.

\spacetune{\enlargethispage{\baselineskip}}
The types and semantics of the constructs used in this paper are illustrated in Figure~\ref{fig:futhark-soacs}: \code{iota} creates an iteration space and \code{zip} and \code{unzip} convert between the structure-of-array (SoA) and array-of-structure (AoS) representations. \code{map} applies a function to each array element, and inclusive \code{scan} computes all the prefixes of the input array with an {\em associative} operator $\odot$, which has neutral element $ne_\odot$ (exclusive scan has $ne_\odot$ as first element). \code{map2}, \code{map3} are like \code{map}, but apply the function to corresponding elements from two and three input arrays, respectively, e.g.,
\noindent
\begin{lstlisting}[numbers=none,language=futhark,mathescape=true,xleftmargin=0pt]
 def map2 f as bs = map ($\lambda$(a, b) -> f a b) <| zip a b
\end{lstlisting}
\vspace{-1ex}
A do-loop expression declares a loop counter $i$ taking values in $0\ldots e^n~\text{-}~1$ and a tuple of loop-variant variables $\overline{x}$ and their initializing expressions $\overline{x}^{init}$; the result of the body expression $e^{body}(i,\overline{x})$ is bound to $\overline{x}$ for the next iteration. Since shadowing of let bindings is allowed, if $\overline{x}$ exists in the scope, it serves as implicit initialization for the syntactic sugar notation 
$\textbf{loop}~\overline{x}~\textbf{for}~i < n~\textbf{do}~ \ldots$   

In-place updates are supported by an uniqueness type technique~\cite{uniqueness-clean} using the syntax $\textbf{let}~x'\text{=}~x~\textbf{with}~[\overline{e}^{ind}] ~\text{=} e^{val}$ that also accepts the syntactic sugar $\textbf{let}~x[\overline{e}^{ind}] = e^{val}$.

\code{reduce\_by\_index\_stream} is essentially a principled, type-safe abstraction~\cite{ad-sc22} of generalized reduction~\cite{log-infer-dep-an}: it is similar to \code{map (f dest) bs}, except that $f$ can only access its first argument (initialized to \code{dest}) by means of the \code{write} function defined above. The latter (atomically) updates the value of some index \code{i} with the value obtained by applying the {\em associative} and {\em commutative} operator $\odot$ to the existing value at \code{i} and the new value \code{v}. 

{\em Please note that up to this point all constructs guarantee correct-by-construction parallelism}, i.e., data races are not possible. The \code{scatter\_stream} \soac{} is similar, except that it overwrites the value at that index instead of accumulating to it (since $\odot$ is not passed as argument). Similar to \code{scatter}, its use is unsafe---WAW dependencies are possible---and dynamic verification is prohibitively expensive; however recent work~\cite{array-props} has presented promising results for static verification of common cases of scatter.

The bottom of Figure~\ref{fig:futhark-soacs} demonstrates the use of loops and \code{scatter\_stream} to flatten and shift the elements of a 2D array by \code{n} positions, where \code{n} is also allowed to be negative: the 1D result array of length \code{m*q} is initialized with zero and each of the \code{m} threads (\code{tid}) executes the loop that writes each of its $q$ elements (\code{xs[tid]}) at the final (shifted) position. Please note the use of sized-dependent types~\cite{size-dep-tps-2,slice-bounds-check}; they are realized by syntactic matching, e.g., passing \code{replicate (q*m) 0} as first argument to \code{scatter\_stream} would result in a type error, because 
\code{[m*q]uint} is syntactically different than 
\code{[q*m]uint}. 

This can be remedied by runtime~verified casting \code{let x'= x :> [m][q]uint}. Sized types are useful, e.g., in specifying that efficient multiplication requires an even sequentialization factor \code{[m][2*q]uint}. 
Finally, Futhark provides a simple FFI to mainstream languages~\cite{futhark-python}, but without support for parameterized components~\cite{alma:ISSAC,mapal_synasc}.\smallskip

\enlargethispage{\baselineskip}

{\em Futhark's optimizing compiler} is primarily aimed at GPU execution. The key code transformation is to map an arbitrary number of levels of application parallelism to the fixed one supported by the hardware.
This is achieved by a technique named {\em incremental flattening}~\cite{futhark-ppopp,futhark-tuning} that systematically applies map fission and map-loop interchange to create perfect parallel nests, and generates multiple code versions, which utilize incrementally more levels of application parallelism. These code versions are composed by guarding them with predicates that compare their degree of utilized parallelism to a threshold, which is subject to autotuning. Importantly, some of the created kernels, named {\em intragroup}, map 
inner parallelism to the \cuda{} block level, such that all intermediate arrays are allocated in shared memory, which offers better bandwidth and latency than global memory. In this case, the guard is augmented with a test that succeeds when the shared-memory footprint of the kernel fits the hardware; otherwise a different code version is chosen. 

A final relevant optimization is {\em{}memory merging}~\cite{futhark-mem-sc22,philip-thesis}, which refers to applying a register-allocation-like algorithm to optimize the reuse of shared-memory buffers, rather than scalars. This is key to enabling large intragroup kernels, since its shared memory must be allocated before running the kernel and a pure setting dictates that arrays are commonly freshly created, e.g., by \code{map}, \code{scan}.
Notably, we use \code{opaque} to explicitly prevent consumer-producer fusion and completely disable~horizontal~fusion~\cite{redomap-fusion}, because it hinders the reuse of shared memory. 

\section{Futhark Compiler Improvements}
\label{sec:compiler-improvements}

\begin{figure}
\begin{lstlisting}[language=futhark, mathescape=true]
def demo [n][q] (ass: [n][q]u64) =
 let (reg: [n][q]u64, shm: [n][2]u64) = unzip
    <| #[toregmem(1)] map f ass@\label{ln:reg-annot}@
  ...
 let x=map($\lambda$tid -> let z=reg[tid] in g z)(iota n)@\label{ln:safe-use1}@
 let yss = map ($\lambda$ tid -> 
    let y= if tid==0 then 0 else shm[tid-1,0] in@\label{ln:unsafe-use1}@
    loop y for i<q do y*y + reg[tid,i]) (iota n)@\label{ln:safe-use2}@
  ...
def main[m][n][q] (A: [m][n][q]u64) = map demo A@\label{ln:intra-kernel}@
\end{lstlisting}
\vspace{-2ex}
\caption{Register Demo: \code{reg} is mapped to registers, \code{shm} is {\em not}.}
\label{lst:reg-optim}
\vspace{-1\baselineskip}
\end{figure}

The intragroup kernels created by incremental flattening allocate all intermediates in shared memory because this is always safe, i.e., accessible by any thread. However, this is inefficient for several reasons: (1) registers have better bandwidth and latency than shared memory, and allow graceful degradation by memory spilling, 
(2) fast memory is scarce, hence both shared {\em and} register memory should be used to maximize performance, (3) Futhark performs deep copies when merging arrays across branches and loop iterations--exacerbating shared-memory footprint--possibly resulting in mutually aliased buffers that are not reused. 

To remedy this, we have designed a compiler pass that allocates arrays in register memory according to user annotations whenever the analysis can verify that this is safe. The analysis consists of a forward and backward pass that execute in sync during a program traversal: the forward pass extends the context with information referring to scalar expansion, array indirection/slicing, and importantly, with the arrays that are targeted to register allocation. The backward pass checks in a conservative manner that the use (read) of these arrays is compatible with register mapping, namely that (i) the inner-parallel sub-kernel that defined the array has the same parallel dimensions as the (current) one that reads it, and (ii) the outermost indices of the read array correspond to the current sub-kernel indices.  If any read violates this pattern, then the array is {\em not} subject to register allocation. 

Other approaches concentrate primarily on improving traffic to shared memory, \textit{e.g.}~\cite{DBLP:conf/iwomp/TalaashrafiMD22},  in particular through the \mbox{OpenACC \cite{OpenACC-compiler}} or OpenMP~\cite{openmp} frameworks.

\enlargethispage{\baselineskip}

Figure~\ref{lst:reg-optim} shows an example: the \code{map} at line~\ref{ln:intra-kernel} forms an intragroup kernel, whose parallel sub-kernels correspond to the \code{map} operations in function \code{demo}. The first (line~\ref{ln:reg-annot}) advises the compiler to try to allocate results in registers (\code{\#[toregmem(1)]}, where \code{1} denotes the parallel levels).  The backward pass rejects \code{shm} because its outer index is \code{tid-1} instead of \code{tid} at line~\ref{ln:unsafe-use1}. The analysis succeeds for \code{reg}, which is indexed with \code{tid} %
at lines~\ref{ln:safe-use2}~and~\ref{ln:safe-use1}.
{Of note, the analysis ``remembers'' that \code{z} is the subarray \code{reg[tid]} but any use of \code{z} in \code{g} still conforms to the required pattern.}

Ifs and loops have recursive bodies and receive special treatment. A loop variant is mapped to registers if (i) its initializer was meant to be in registers and (ii) its shape is invariant through the loop. If its use in the loop body is not amenable to register mapping, it is copied to shared memory at the start and loaded to registers at the end---this treatment eliminates the deep-copy and aliasing issues of shared memory. Ifs are treated similarly, but their result \code{r} needs to be refined through \code{\#[inform\_pardim\_only(k)] manifest r}, so as to inform the parallel rank \code{k}. In principle, it seems possible to aggressively map arrays to registers without any annotations.

More advanced analyses than ours were used to disambiguate challenging indexing patterns statically~\cite{PolyPluto1,venkat:pldi15,civ-an} or dynamically~\cite{R-LRPD,jaycos:DTLS}.
Our analysis treats simpler accesses, which still cover well the common case of Futhark programs. Its scope is not limited to integer arithmetic; in fact, it would significantly accelerate a number of real-world applications written in Futhark~\cite{stl-climate,finance1,app-kd-tree}, and enable efficient radix sort~\cite{cub_cccl} and Flash Attention~\cite{flash-attention}.

\section{Big Integer Arithmetic in Futhark}
\label{sec:futhark-impls}

This section discusses Futhark implementations for addition, multiplication, and division. We aim to highlight clean, high-level, hardware-neutral, and pure specifications that 
\begin{enumerate*}
\item[(1)] are constructed by composing and nesting SOACs, such as map and scan, and guarantee (for most parts) correct-by-construction parallelism,
\item[(2)] provide good support for abstraction, such as size-dependent types and higher-order functions, and
\item[(3)] permit optimization of memory placement by register-scheduling directives--such as \lstinline{#[toregmem(1)]} applied to a map--but are otherwise memory agnostic, thus freeing the programmer from the tedious and challenging task of managing memory.
\end{enumerate*}
In comparison, \cuda{} implementations (1) do not guarantee the absence of data races, (2) require manually splitting the application into {\sc cpu} host code and a number of {\sc gpu} kernels, and (3) require careful allocation and reuse of arrays into/from the most suited memory level (global, shared, private).  

\subsection{Addition}
\label{subsec:futhark-badd}

The key idea we build on is that data-parallel addition can be expressed as a map-scan-map composition~\cite{blelloch-scan,OanceaWatt2025}. Assuming some digit type {\tt uint} and two $n$-digit unsigned integers $x$ and $y$, their addition $x + y$ can be computed as in Figure~\ref{lst:golden-add}.
In essence, {\tt map2 $\ominus_{1}$ x y} computes whether the per-digit addition overflows or results in the maximal {\tt uint} value. The result (an array of boolean tuples) is passed to an exclusive prefix sum ({\tt scan$^{exc} ~\odot~$(false,true)}) that propagates the carry for each digit, and the last map applies the corresponding carry to the per-digit addition.

Figure~\ref{fig:badd-F-code} shows the optimized code for addition.
Function {\tt badd} is intended to compute $ipb$ instances of $n\cdot q$ digit additions within a \cuda{} block of $ipb\cdot n$ threads, i.e., $q$ is the sequentialization factor. 
This is hinted by the dependently-sized type {\tt[$ipb* n$][$q$]uint} for the array arguments and result at lines~\ref{ln:arrtp1}~and~\ref{ln:arrtp2}.
Since arguments $ass$ and $bss$ are assumed elements of a batch computation---e.g., \code{map2 badd asss bsss}---they are first copied to register memory at lines~\ref{ln:manifest1}-\ref{ln:manifest2}, but only if they reside in global memory, as indicated by the {\tt\#[glb2reg\_only(1)]} annotation. The latter might not be the case due to fusion. 

Lastly, computation is discharged to function {\tt baddReg} (line~\ref{ln:baddReg}), which maintains its arguments and results in registers and implements the efficiently~sequentialized version of the code in Figure~\ref{lst:golden-add}: $f1$ (lines~\ref{ln:f1beg}-\ref{ln:f1end}) corresponds to $\ominus_{1}$, except that (1) it processes sequentially $q$ consecutive elements and (2) it returns the per-digit addition and carry results (array {\tt rcs}) together with one carry per-thread, denoting the carry propagation across its $q$ elements.

\begin{figure}
\begin{lstlisting}[language=futhark, mathescape=true]
def $\ominus_{1}$ a b = (a+b > a, a+b == uint.highest)
def $\odot$ (o1,h1) (o2,h2) = ((o1&&h2)||o2, h1&&h2) 
def $\ominus_{3}$ a b (c, _) = a + b + bool2uint c
def badd $x~ y$ = map3 $\ominus_{3}~ x~ y$ <|
          scan$^{exc}$ $\odot$ (false,true) <| map2 $\ominus_{1}~ x~ y$ 
\end{lstlisting}
\vspace{-2ex}
\caption{Golden Data-Parallel Addition. The operator \lstinline{<|} pipes the result of a computation to the last input of the next computation.}
\label{lst:golden-add}
\vspace{-\baselineskip}
\end{figure}
\begin{figure}
\begin{lstlisting}[language=futhark]
def carryBop (c1: cT) (c2: cT) =
  if (c2 & 4) != 0 then c2
  else let res = ( (c1 & (c2 >> 1)) | c2 ) & 1
       in  res | (c1 & c2 & 2) | (c1 & 4)
-- input and result are placed in registers
def baddReg [ipb][n][q] (ass: [ipb*n][q]uint)@\label{ln:baddReg}@
     (bss: [ipb*n][q]uint) : [ipb*n][q]uint =@\label{ln:arrtp2}@
  let f1 as bs tid : (cT, [q](uint, cT)) =@\label{ln:f1beg}@
    let (carry,rcs) = (2, replicate q (0, 2))
    let seg_start = tid % n == 0 in
    loop (carry, rcs) for i < q do
      let (a, b) = (as[i], bs[i])
      let r = a + b
      let c = bool2cT (r < a)
      let c1= (bool2cT(r == highest_uint))<< 1
      let c2= (bool2cT(i==0 && seg_start))<< 2
      let c = c | c1 | c2
      let carry = carryBop carry c
      let rcs[i] = (r,c)
      in  (carry, rcs)@\label{ln:f1end}@
  let (carry_thds,rcss) = opaque <| unzip <|
   #[toregmem(1)] map3 f1 ass bss (iota (ipb*n))
  -- scan across threads, neutral element is 2
  let carry_thds = scan carryBop 2 carry_thds@\label{ln:scan}@
  -- adjust result with thread prefix
  let f2 rcs tid : [q]uint =@\label{ln:f2beg}@
    let (rs, cs) = unzip rcs
    let seg_start = tid % n == 0
    let carry = if seg_start then 2
                else carry_thds[tid-1]
    let rs' = #[scratch] replicate q 0 in
    (loop (rs', carry) for i < q do
       let rs'[i]= rs[i] + cT2uint (carry & 1)
       in (rs', carryBop carry cs[i])).0@\label{ln:f2end}@
  in  opaque <| unzip <| #[toregmem(1)]
      map2 f2 rcss (iota (ipb*n))
-- ass and bss may reside in global memory
def badd [ipb][n][q] (ass : [ipb*n][q]uint) 
    (bss : [ipb*n][q]uint) : [ipb*n][q]uint =@\label{ln:arrtp1}@
  let areg = #[glb2reg_only(1)] manifest ass@\label{ln:manifest1}@
  let breg = #[glb2reg_only(1)] manifest bss@\label{ln:manifest2}@
  in  baddReg areg breg
\end{lstlisting}\vspace{-2ex}
\caption{Futhark efficiently-sequentialized code for addition. }
\label{fig:badd-F-code}
\vspace{-1\baselineskip}
\end{figure} 
Line~\ref{ln:scan} computes the prefix sum of the per-thread carry results. Function \texttt{carryBop} has semantics similar to $\odot$, except that (1) it encodes a boolean tuple in the first two least-significant bits of an integer (of some type {\tt cT}), and (2) it lifts $\odot$ to operate across addition instances by setting the third bit at the start of each new instance. Finally, $f2$ (lines~\ref{ln:f2beg}-\ref{ln:f2end}) corresponds to $\ominus_{3}$: it computes and applies the final carry to each of the $q$ elements owned by a thread.

Please note that the code of {\tt baddReg} allows the arguments {\tt ass} and {\tt bss} and the intermediate array {\tt rcss} to be safely allocated in registers, because accesses do not overlap across threads---e.g., the producer and consumer maps ($f1$ and $f2$) write and read  element {\tt rcss[tid]} only with thread {\tt tid} (i.e., in the iterations numbered {\tt tid}). 

Finally, we remark that while the code is hardware neutral, memory agnostic, and relatively high level, it is still significantly more complex, albeit much faster, than the one in Figure~\ref{lst:golden-add}.
Achieving optimal expression {\em and} performance is, in principle, possible by designing and exposing to the user more advanced scheduling primitives~\cite{sched-lang}, which would generate the code of Figure~\ref{fig:badd-F-code} from Figure~\ref{lst:golden-add}.

\subsection{Multiplication}
\label{subsec:futhark-bmul}

\begin{figure}
\vspace{-0.4ex}
\begin{lstlisting}[language=futhark,mathescape=true,numbers=left]
def cpShm2Reg[n][Q]'t (shm:[n*Q]t): *[n][Q]t =@\label{ln:shm2reg}@
  let ff tid = #[sequential]
    map ($\lambda$q-> shm[tid*Q+q]) (iota Q)
  in  #[toregmem(1)] map ff (iota n)
  
def cpReg2Shm[n][Q]'t x (reg:[n][Q]t):*[n*Q]t=@\label{ln:reg2shm}@
  let shm = #[scratch] replicate (n*Q) x
  let f (A: *acc([n*Q]t)) tid : acc([n*Q]t)=
    loop A for q < Q do
      write A (tid*Q + q) (reg[tid, q])
  in scatter_stream shm f (iota n)

def halfConv [n] (Q: i64) (off: i64) (k: i64) 
    (ash: [n]uint)(bsh: [n]uint) : [Q+2]uint =
  let cs = #[sequential] replicate Q (0, 0, 0)
  let cs = loop cs for i < k + 1 do@\label{ln:loop1-beg}@
        let j = k - i in
        loop cs for q < Q do@\label{ln-perf-critical-loop}@
          cs with [q] = oneConvMul off i (j+q)
                                 ash bsh cs[q]@\label{ln:loop1-end}@
  let cs = loop cs for qm1 < Q-1 do@\label{ln:loop2-beg}@
        let q = qm1 + 1 in
        loop cs for i < Q-q do
          cs with [i+q]=oneConvMul off (k+q) i
                               ash bsh cs[i+q]@\label{ln:loop2-end}@
  in combineQ cs@\label{ln:combine}@
  
def threadConv [x] (Q:i64) (n:i64) (ash:[]uint)
 (bsh:[]uint) (tid:i64): ([Q+2]uint,[Q+2]uint)=
  let offset = n * (2 * Q) * (tid / n)
  let k1 = Q * (tid % n)
  let lhcs0 = halfConv Q offset k1 ash bsh@\label{ln:thd-conv-fwd}@
  let k2 = n * (2*Q) - i32.i64 Q - k1
  let lhcs1 = halfConv Q offset k2 ash bsh@\label{ln:thd-conv-bwd}@
  in  (lhcs0, lhcs1)

def convShmQ[s] (n:i64) (Q:i64) (Ash:*[s]uint)@\label{ln:conv-beg}@
        (Bsh: *[s]uint) : ([s]uint, [s]uint) =
  let (lhcss0,lhcss1)= unzip <| #[toregmem(1)]@\label{ln:conv-res}@
    map (threadConv Q n Ash Bsh)(iota (ipb*n))@\label{ln:map-conv}@
  in  manifestShm Ash Bsh lhcss0 lhcss1@\label{ln:conv-end}@

def bmulReg[ipb][n][Q] (Areg:[ipb*n][2*Q]uint)@\label{ln:bmulreg}@
    (Breg: [ipb*n][2*Q]uint): [ipb*n][2*Q]uint=
  let (Lsh,Hsh)=convShmQ n Q (cpReg2Shm 0 Areg)
                             (cpReg2Shm 0 Breg)
  in  baddReg (cpShm2Reg Lsh) (cpShm2Reg Hsh)@\label{ln:adding-lhcs}@
\end{lstlisting}~
\begin{center}
\includegraphics[width=0.4\textwidth]{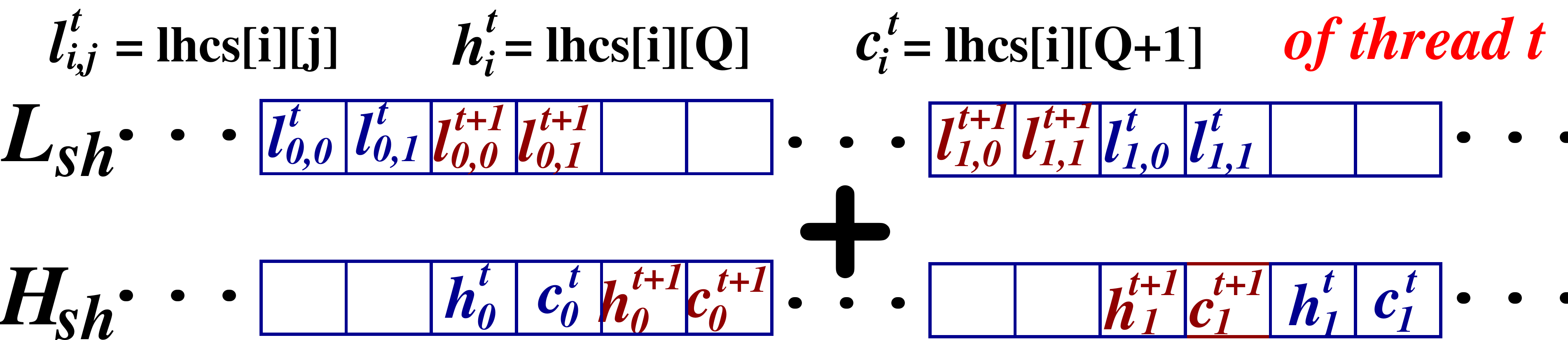}
\end{center}
\caption{Futhark code for multiplication. The diagram illustrates the placement of thread-private (register) array {\tt lhcs: [2][Q+2]uint $\cong$ (lhcs0,lhcs1)} in shared-memory arrays {\tt Lsh} and {\tt Hsh}, which is achieved by the call to {\tt manifestShm} on line~\ref{ln:conv-end}. {\tt Lsh} and {\tt Hsh} are then loaded to registers and added at line~\ref{ln:adding-lhcs} to complete the algorithm.  
}
\vspace{-3ex}
\label{fig:bmul-F-code}
\end{figure}

Assuming two $m$-digit integers $A$ and $B$ the quadratic multiplication algorithm is summarized by the formula:
\[
C_{k} = \sum_{\substack{i+j = k \\0\leq i,j,k < m}} A_i \cdot B_j
\]
which, however, does not handle overflow across $C$'s digits.

Figure~\ref{fig:bmul-F-code} highlights the code structure of multiplication in Futhark, which follows the strategy proposed in~\cite{OanceaWatt2025}:
Function {\tt bmulReg} (line~\ref{ln:bmulreg}) is~intended~to receive two register-allocated arguments and also produce their multiplication result in register memory. The implementation consists of (i) performing the quadratic convolution ({\tt convShmQ}), which has arguments and results allocated in shared memory, and (ii) adding the results of convolution ({\tt baddReg}). The conversion between shared and register memory is explicitly performed by the functions {\tt cpShm2Reg} and {\tt cpReg2Shm}, defined at lines~\ref{ln:shm2reg}~and~\ref{ln:reg2shm}. The former uses a sequentialized map to load thread elements from shared memory 
and the latter uses {\tt scatter\_stream}, by which each thread writes its $Q$ registers to consecutive positions in a fresh and flat shared-memory buffer. 

Function {\tt convShmQ} (lines~\ref{ln:conv-beg}-\ref{ln:conv-end}) performs the convolution by mapping  {\tt threadConv} to each of the $ipb\cdot n$ threads (line~\ref{ln:map-conv}). Assuming for simplicity $ipb=1$, each thread $tid$ sequentially computes $2\cdot Q$ elements of the result in two steps: (1) $Q$ elements in forward direction, i.e., at indices $tid\cdot Q ~\ldots~ (tid+1)\cdot Q - 1$ (line~\ref{ln:thd-conv-fwd}), and (2) their $Q$ symmetric-opposite elements, at indices $n\cdot2\cdot Q - (tid+1)\cdot Q ~\ldots~ n\cdot2\cdot Q - tid\cdot Q - 1$ (line~\ref{ln:thd-conv-bwd}). 

This scheduling {\em requires an even sequentialization factor in order to enforce a perfectly balanced workload across threads, in which each thread computes $n\cdot Q$ scalar multiplications}, e.g., the first and last result elements require $1$ and $n\cdot Q-1$ multiplications, respectively, the second and penultimate require $2$ and $n\cdot Q - 2$, and so on.  

Function {\tt halfConv} computes $Q$ consecutive elements of the result, i.e., half of a thread's work. The first loop nest (lines~\ref{ln:loop1-beg}-\ref{ln:loop1-end}) performs a workload equal to that of the first element for all $Q$ elements, and the second loop nest (lines~\ref{ln:loop2-beg}-\ref{ln:loop2-end}) performs the missed work for the last $Q-1$ result elements. This structure allows efficient unrolling and scalarization of the inner loop of count $Q$ (line~\ref{ln-perf-critical-loop}) and of the second loop nest, which is performance critical.

The call {\tt oneConvMul o i j A B cs[q]} performs one scalar multiplication $A_{o+i} \cdot B_{o+j}$ that returns a triplet of low, high, and carry digits, which are accumulated to the current result stored in {\tt cs[q]}, hence {\tt cs: [Q][3]uint}. 
Importantly, Futhark does not support $128$-bit arithmetic: when {\tt uint} has $64$ bits we simulate it by performing four $64$-bit multiplications, which is less efficient than a \cuda{} $128$-bit multiplication, yielding an application-level slowdown of about $1.33\times$ on highest multi-precision sizes.

After all $Q$ consecutive elements of the result are fully computed, they are aggregated into $Q$ digits plus a high and a carry digit by the call to {\tt combineQ} on line~\ref{ln:combine},~hence the result of a thread's half convolution has type {\tt[Q+2]uint}.
It follows that the half-convolution results across all threads, denoted {\tt lhcss0} and {\tt lhcss1} on line~\ref{ln:conv-res}, have types {\tt[n][Q+2]} and are stored in registers ({\tt\#[toregmem(1)]}).
Finally, the call to {\tt manifestShm} on line~\ref{ln:conv-end} places the {\tt lhcss0} and {\tt lhcss1} arrays in shared-memory arrays {\tt Lsh} and {\tt Hsh}, as illustrated by the bottom picture of Figure~\ref{fig:bmul-F-code} and then {\tt Lsh} and {\tt Hsh} are loaded to registers and added at line~\ref{ln:adding-lhcs} to complete the algorithm.  

\vbox{
\subsection{Division}
\label{subsec:futhark-bdiv}

The Futhark implementation draws inspiration from the \cuda{} parallelization strategy presented in~\cite{OanceaWatt2026a} that implements Watt's algorithm~\cite{watt2023}. The latter  allows the computation to be carried out entirely in the integral domain and requires at least five full-precision multiplications to compute the quotient and remainder. For more algorithmic detail see the original paper~\cite{watt2023}.
}
\begin{figure}
\begin{lstlisting}[language=futhark]
def bmulG [n][Q] (Q': i64) (m: i64)@\label{ln:bmulG-beg}@
                 (Ash: *[(1*n)*(2*Q)]uint)
                 (Bsh: *[(1*n)*(2*Q)]uint) =
  let (Lsh,Hsh) = convShmQ n Q' Ash Bsh in
  (cpShm2RegPad m 0 Lsh, cpShm2RegPad m 0 Hsh)
  
-- computes the first m digits of A*B
def varPrecMul[Q][n] (m:i64)(Areg: [n][2*Q]uint)@\label{ln:var-mul-beg}@
           (Breg: [n][2*Q]uint) : [n][2*Q]uint =
  let Ash= cpReg2Shm (Areg :> [1*n][2*Q]uint)
  let Bsh= cpReg2Shm (Breg :> [1*n][2*Q]uint)
  let Q' = if m <= 1*n then 0 else
           if m <= 2*n then 1 else
           if m <= 4*n then 2 else Q
  let (Lreg,Hreg) = 
   #[inform_pardim_only(1)] manifest
   (match Q'--bmul1 is specialized to compute
    case 0 -> bmul1 m Ash Bsh--one res/thread@\label{ln:bmul0}@
    case 1 -> bmulG 1 m Ash Bsh -- uses Q'= 1@\label{ln:bmul1}@
    case 2 -> bmulG 2 m Ash Bsh -- uses Q'= 2@\label{ln:bmul2}@
    case _ -> bmulG Q m Ash Bsh)-- uses Q'= Q@\label{ln:bmul4}@
  in (baddReg Lreg Hreg) :> [n][2*Q]uint@\label{ln:var-mul-end}@
  
-- computes max non-zero index of xss plus 1
def prec [n][q] (xss : [n][q]uint) : i32 =@\label{ln:prec-beg}@
  let f (A: *acc([1]i32)) tid : acc([1]i32) =
    let ind = loop ind = 0i32 for i < q do@\label{ln:loop-max-beg}@
                if xss[tid, i] == 0 then ind
                else i32.i64 (q * tid + i + 1)@\label{ln:loop-max-end}@
    in if ind > 0 then write A 0 ind else A@\label{ln:histo-write}@
  in (reduce_by_index_stream (replicate 1 0i32)@\label{ln:histo}@
                     (i32.max) 0 f (iota n))[0]@\label{ln:prec-end}@
                     
def step [n][q] (h: i32) (m: i32) (l: i32)
         (vs: [n][2*q]uint, pv: i32)
         (ws: [n][2*q]uint) : [n][2*q]uint =
 let pw = prec ws@\label{ln:prec1}@
 let (sgn,xs)= powDiff(h-m)(l-2)(vs,pv)(ws,pw)
 let ys = varPrecMul(pw+prec xs) ws xs-- ws*xs@\label{ln:step-mul}@
 let ws_sft = shift m ws        -- ws << m@\label{ln:shift1}@
 let ys_sft = shift (2*m - h) ys-- ys << 2*m-h@\label{ln:shift2}@
 in  #[inform_pardim_only(1)] manifest <|
   if ( sgn == 1 )
   then baddReg' ws_sft ys_sft --ws_sft+ys_sft@\label{ln:add0}@
   else let isZero = nullUpToInd (2*m - h) ys@\label{ln:null2ind}@
        let ys_sft'= if isZero then ys_sft
                     else baddOne ys_sft -- +1@\label{ln:add1}@
        let ys_sft'= #[inform_pardim_only(1)]
                     manifest ys_sft' in
        bsubReg' ws_sft ys_sft'--ws_sft-ys_sft@\label{ln:sub0}@
\end{lstlisting}
\caption{Futhark code used to implement the whole shifted inverse.}
\label{fig:bdiv-F-code}
\spacetune{\vspace{0.15\baselineskip}}
\end{figure}

Figure~\ref{fig:bdiv-F-code} is intended to give the gist of the Futhark implementation by presenting incomplete code corresponding to the {\tt step} function that is run in a loop and consists of operations on integers whose precision varies through the loop. Since the algorithm cost is dominated by multiplication, we follow the strategy used in~\cite{OanceaWatt2026a} that specializes the multiplications to the actual precision of the input and ``pads'' the linear operations to full precision, i.e., the declared precision of division's input. Examples of the latter include shifting (line~\ref{ln:shift1},\ref{ln:shift2}), adding (lines~\ref{ln:add0},\ref{ln:add1}), subtracting (lines~\ref{ln:sub0}), and testing if all least-significant digits up to an index are null (line~
\ref{ln:null2ind}). As a sample, the implementation of {\tt prec}, which computes the precision of a given integer, is shown between lines~\ref{ln:prec-beg}-\ref{ln:prec-end}: it applies the {\tt reduce\_by\_index\_stream} \soac{} (line~\ref{ln:histo}) that uses atomic updates to compute the maximal non-zero index across the per-thread results. The latter are independently aggregated over the {\tt2*Q} elements of each thread in the loop at lines~\ref{ln:loop-max-beg}-\ref{ln:loop-max-end}; a thread performs {\em an atomic update} only when it owns a non-zero element (using {\tt write} at line~\ref{ln:histo-write}).

Multiplication in variable precision is implemented by the {\tt varPrecMul} function (lines~\ref{ln:var-mul-beg}-\ref{ln:var-mul-end}), called on line~\ref{ln:step-mul} and also in the previous line inside {\tt powDiff} (not shown).
The key idea is to {\em specialize the computation to the required precision of the result}, denoted by the $m$ argument, {\em by decreasing accordingly the value of the half-sequentialization factor {\tt Q'}}: In the case in which the result precision $m$ is less than or equal to the number of threads $n$, we use a specialized implementation of multiplication---the {\tt bmul1} call on line~\ref{ln:bmul0}---that computes an element of the result with each thread, albeit in a thread-unbalanced way. 

Otherwise, we use the value of {\tt Q' $\in \{ 1, 2, 4 \}$} that best matches the desired precision---see the calls to {\tt bmulG} at lines~\ref{ln:bmul1}, \ref{ln:bmul2}, \ref{ln:bmul4}. The function {\tt bmulG} (i) simply calls {\tt convShmQ}---shown in Figure~\ref{fig:bmul-F-code} and discussed in Section~\ref{subsec:futhark-bmul}---with the reduced sequentialization factor {\tt Q'} and (2) loads the results back in register memory. The latter is accomplished by function {\tt cpShm2RegPad}, which is similar to {\tt cpShm2Reg} except that it sets the indices greater or equal to its first argument ($m$) with a pad value ($0$). This is necessary because certain multiplications (in {\tt powDiff}) are performed modulo the $m$-th power of the base, which requires zeroing out said digits of the result.   

\newpage
\section{Evaluation}
\label{sec:evaluation}

\begin{table*}[t]
    \caption{Performance of Addition in \textbf{GB/sec}. The number of bytes accessed from global memory for both \textbf{1-Add} and \textbf{6-Add} is:  $3\cdot\textbf{NumInsts}\cdot\textbf{NumBits}/8$, i.e., each digit of the two inputs and result is accessed once. A100's peak bandwidth is \textbf{1555} GB/s.}
    \label{tab-add}\vspace{-4ex}
\begin{center}
\begin{tabular}{|c|c||r|c|c|c||r|c|c|c|}
    \hline
    \textbf{Num} & \textbf{Num} & \textbf{1-Add} & \textbf{1-Add} & \textbf{1-Add} & \textbf{1-Add} & \textbf{6-Add} & \textbf{6-Add} & \textbf{6-Add} & \textbf{6-Add} \\
    \textbf{Bits}& \textbf{Insts}& \textbf{CGBN}  & \textbf{CudaP} & \textbf{F-Reg} & \textbf{F-Shm} & \textbf{CGBN}  & \textbf{CudaP} & \textbf{F-Reg} & \textbf{F-Shm}\\
    \hline\hline
    $2^{18}$ & $2^{14}$ & 369  &  1046 & 1303  &  922 & 362   &  575 &  552 & \textcolor{red}{---}\\\hline 
    $2^{17}$ & $2^{15}$ & 368  &  1209 & 1348  & 1298 & 353   &  826 &  648 & 277\\\hline 
    $2^{16}$ & $2^{16}$ & 376  &  1358 & 1338  & 1219 & 353   &  861 &  693 & 393\\\hline
    $2^{15}$ & $2^{17}$ & 329  &  1363 & 1341  & 1344 & 321   &  848 &  675 & 407\\\hline
    $2^{14}$ & $2^{18}$ & 581  &  1334 & 1342  & 1342 & 546   &  875 &  704 & 405\\\hline
    $2^{13}$ & $2^{19}$ & 1238 &  1350 & 1341  & 1343 & 1207  &  881 &  705 & 406\\\hline
    $2^{12}$ & $2^{20}$ & 1329 &  1359 & 1337  & 1338 & 1189  &  882 &  711 & 405\\\hline
\end{tabular}
\end{center}
\end{table*}

\begin{table*}[t]
    \caption{Performance of Multiplication
    on one multiplication and a polynomial involving four multiplications and three additions.
    The \textbf{CudaP} columns report the runtime in $ms$ of the CUDA implementation from~\cite{OanceaWatt2025}. The columns denoted \textbf{CGBN}, \textbf{F-Reg} and \textbf{F-Shm} show the slowdown {\em vs.} \textbf{CudaP} of the \cgbn{} library, and the Futhark code that is compiled with and without support of allocating intermediate buffers in register memory.}
    \label{tab-mul}\vspace{-3ex}
\begin{center}
\begin{tabular}{|c|c||c|l|l|l||c|l|l|l|}
    \hline
    \textbf{Num} & \textbf{Num} & \textbf{1-Mul} & \textbf{1-Mul} & \textbf{1-Mul} & \textbf{1-Mul} & \textbf{Poly} & \textbf{Poly} & \textbf{Poly} & \textbf{Poly} \\\hline
    \textbf{Bits}& \textbf{Insts}& \textbf{CudaP} & \textbf{CGBN /} & \textbf{F-Reg /} & \textbf{F-Shm /} & \textbf{CudaP} & \textbf{CGBN /} & \textbf{F-Reg /} & \textbf{F-Shm /}\\
    & & $ms$ & \textbf{CudaP} & \textbf{CudaP} & \textbf{CudaP} & $ms$ & \textbf{CudaP} & \textbf{CudaP} & \textbf{CudaP}\\
    \hline\hline
    $2^{19}$ & $2^{13}$ & 420.4   &  ---  & 1.35 &  --- & 1769.9 &  --- & 1.28 & ---\\\hline    
    $2^{18}$ & $2^{14}$ & 207.3 &  35.1 & 1.38  & 1.45 & 855.5 &  49.9 & 1.34 & ---\\\hline 
    $2^{17}$ & $2^{15}$ & 105.5 &  4.48 & 1.37  & 1.41 & 435.0 &  21.0 & 1.35 & 1.38\\\hline 
    $2^{16}$ & $2^{16}$ & 57.0  &  1.21 & 1.31  & 1.31 & 225.1 &  1.22 & 1.33 & 1.34\\\hline
    $2^{15}$ & $2^{17}$ & 30.6  &  1.16 & 1.30  & 1.30 & 119.6 &  1.00 & 1.34 & 1.35\\\hline
    $2^{14}$ & $2^{18}$ & 19.3  &  1.01 & 1.16  & 1.16 & 66.9  &  0.91 & 1.35 & 1.35\\\hline
    $2^{13}$ & $2^{19}$ & 12.4  &  0.93 & 1.10  & 1.10 & 40.7  &  0.82 & 1.35 & 1.37\\\hline
    $2^{12}$ & $2^{20}$ & 8.9   &  0.91 & 0.82  & 0.82 & 22.1  &  0.79 & 1.36 & 1.40\\\hline
\end{tabular}
\end{center}
\end{table*}
\begin{table*}[t]
    \caption{Performance of Division. Columns $3$, $5$ and $9$ report the runtimes of division for the code version \textbf{CudaP}, \textbf{F-Reg} and \textbf{CGBN}, respectively. Columns denoted \textbf{Div/Mul} denote the factor by which division is slower than multiplication (\textsc{div-ov}) for the code version of the previous column. The other columns report speedup between the specified code versions.}
    \label{tab-div}\vspace{-3ex}
\begin{center}
\begin{tabular}{|c|c||c|r||c|r|r|r||c|r|r|}
    \hline
    \textbf{Num} & \textbf{Num} & \textbf{CudaP} & \textbf{Div /} & \textbf{F-Reg} & \textbf{Div /} & \textbf{F-Reg /} & \textbf{F-Shm /} & \textbf{CGBN} & \textbf{Div /} & \textbf{CudaP /} \\\hline
    \textbf{Bits}& \textbf{Insts}& \textbf{\em ms} & \textbf{Mul\h} & \textbf{\em ms} & \textbf{Mul\h}  & \textbf{CudaP\h}   & \textbf{F-reg\h} & \textbf{\em ms} & \textbf{Mul\h} & \textbf{CGBN\h} \\\hline
    $2^{19}$ & $2^{13}$ & ---   &  ---  & 3166.4 & 5.58 & ---  &  ---  &  --- &  --- &  ---\\\hline    
    $2^{18}$ & $2^{14}$ & 1404.3&  6.77 & 1626.9 & 5.70 & 1.16 &  ---  &  --- &  --- &  ---\\\hline 
    $2^{17}$ & $2^{15}$ & 702.1 &  6.66 &  962.0 & 6.66 & 1.37 &  ---  &  --- &  --- &  ---\\\hline 
    $2^{16}$ & $2^{16}$ & 395.2 &  6.94 &  606.0 & 8.10 & 1.53 &  ---  &  --- &  --- &  ---\\\hline
    $2^{15}$ & $2^{17}$ & 297.9 &  9.73 &  312.2 & 7.82 & 1.05 &  2.37 & 82.0 & 2.31 & 3.64\\\hline
    $2^{14}$ & $2^{18}$ & 284.2 & 14.74 &  278.7 &12.49 & 0.98 &  1.31 & 44.3 & 2.27 & 6.42\\\hline
    $2^{13}$ & $2^{19}$ & 191.2 & 15.46 &  254.5 &18.75 & 1.33 &  1.42 & 26.0 & 2.26 & 7.36\\\hline
\end{tabular}
\end{center}
\end{table*}

\paragraph{Hardware, Code Versions, Datasets, Availability} The evaluation was run under the Red Hat Enterprise Linux 8.10 operating system on an NVIDIA A100 GPU, which has $6912$ cores, $1555$ GB/sec peak global-memory bandwidth and 19.5 TFLOP/s \texttt{FP32} peak performance.

\text{CudaP} denotes our earlier \cuda{} implementations of addition, multiplication, and division reported in~\cite{OanceaWatt2025} and \cite{OanceaWatt2026a}, while \text{CGBN} denotes those using the \text{CGBN} library.
\text{F-Reg} and \text{F-Shm} denote the (same) Futhark implementation reported in this paper, compiled with and without the register-placement optimization, i.e., \text{F-Shm} allocates intermediate arrays exclusively in shared memory.

The datasets are chosen to vary the integer precision in number of bits (\texttt{Num Bits}) from $2^{12}$  to  $2^{19}$ and the number of parallel instances (\texttt{Num Insts}) from  $2^{20}$ down to $2^{13}$, such that $\texttt{Num Bits} \cdot \texttt{Num Insts} = 2^{32}$. The type {\tt uint} is instantiated as a $64$-bit unsigned integer.
Addition and multiplication use random datasets, since the workload is only sensitive to the integer precision, not its value. 

Division borrows the setup used in~\cite{OanceaWatt2026a}: denoting by $M$ the dataset declared precision in {\tt uint} words, the actual precision of the dividend is set to $M - 2$ and the precision of the divisor is randomly selected between $2$ and $M/2$. The rationale is that this configuration always performs the maximal number of iterations in the loop that exhibits precision-variant multiplications.
The code and benchmarking setup will be made 
available at~\cite{FutharkMidIntRepository}.

\spacetune{\enlargethispage{\baselineskip}}
\paragraph{Addition Results}
Table~\ref{tab-add} shows the memory throughput in \textbf{GB/sec} on the two benchmarks examined in~\cite{OanceaWatt2025}: one in which the per-instance work consists of one addition (\textbf{1-Add}) and another in which it consists of six additions (\textbf{6-Add}). For both, the number of bytes accessed from global memory is considered $3\times$ that of one input, i.e., each byte of the two inputs/result must be read/written at least once. The computation of \textbf{6-Add} is fused so that all intermediates are held in fast memory, hence the global-memory accesses are the same as \textbf{1-Add}.

The results in Table~\ref{tab-add} show that \cgbn{} offers poor performance on precisions higher than $2^{13}$ but excellent scalability, in that the throughput of \textbf{6-Add} nearly equals that of \textbf{1-Add}, i.e., six additions are performed at the cost of one. The \textbf{1-Add} performances of \text{CudaP}, \text{F-Reg}, and \text{F-Shm} are close on all but the largest dataset, where \text{F-Reg} is about $1.25\times$ and $1.41\times$ faster than \text{CudaP} and \text{F-Shm}, respectively.
However, on \textbf{6-Add}, \text{CudaP} outperforms \text{F-Reg} with factors between $1.04-1.27\times$. We attribute the slowdown to the Futhark compiler performing (i) suboptimal elimination of redundant barriers, and (ii) internal index calculation in $64$-bit arithmetic, which is costly when the computation is not entirely memory~bound.

Finally, \textbf{6-Add} demonstrates the impact of the register-placement optimization both in terms of (i) scalability: \text{F-Shm} runs out of shared memory on the largest dataset, and (ii) performance: \text{F-Reg} outperforms \text{F-Shm} by $1.66-2.4\times$ factors because registers offer better latency and bandwidth than shared memory. 

\paragraph{Multiplication Results}
Table~\ref{tab-mul} uses \text{CudaP} as a baseline and presents its runtimes and its speedup in comparison to the other three code versions on the two benchmarks reported in~\cite{OanceaWatt2025}: \textbf{1-Mul} denotes $a\cdot b$ and \textbf{Poly} denotes $(a\cdot a + b)\cdot(b\cdot b + b) + a\cdot b$, and executes four multiplications and three additions. Since multiplication uses the quadratic algorithm and the precision is increased by a factor of $2$ while the number of instances is decreased by the same factor, we expect that the runtime increases by a factor of about $\frac{2^2}{2} = 2\times$ with each larger precision.  

Key takeaways are threefold:
{\em First}, \text{CudaP} is faster than \cgbn{} on precisions greater than or equal to $2^{15}$ bits.

{\em Second}, \text{F-Reg} is slower than \text{CudaP} with factors between $0.98-1.53\times$. These are misleading because the main reason for slowdown is that \cuda{} supports efficient $128$-bit multiplications while it remains to be supported in Futhark. Porting Futhark's {\tt oneConvMul} implementation to \textbf{CudaP} slows it down by \text{$\sim$}$1.33\times$ on the larger datasets. This hints that in fact \text{F-Reg} is within $95\%$ of \text{CudaP} performance on all datasets and may even win on \textbf{1-Mul}.

{\em Third}, \text{F-Reg} has about the same performance as \text{F-Shm}, but starts gaining on the larger datasets. As well, \text{F-Shm} cannot run \textbf{Poly} in precision $2^{18}$ and higher. In contrast, \text{F-Reg} can actually run $2^{13}$ instances of  precision $2^{19}$ resulting in {\em scalable} runtimes of $568$ and $2272~ms$ on \textbf{1-Mul} and \textbf{Poly}, respectively. By {\em scalable} we mean that the runtimes above are, as expected, about $2\times$ larger than the ones of running $2^{14}$ instances of precision $2^{18}$.

\spacetune{\enlargethispage{\baselineskip}}
\paragraph{Division Results}
Table~\ref{tab-div} presents the results for one division operation. Columns {\tt$3$-$4$}, {\tt$5$-$6$} and {\tt$9$-$10$} report the runtimes of code versions \text{CudaP}, \text{F-Reg} and \cgbn{} and the factor by which a division is more expensive than a full multiplication of the same code version, abbreviated {\sc div-ov}. The {\sc div-ov} of \text{CudaP} is significantly higher but more accurate than the one reported in~\cite{OanceaWatt2026a} because the latter uses as baseline a slower multiplication.

The numbers for \text{CudaP} and \cgbn{} indicate that \cgbn{} is using a different algorithm for division, since its {\sc div-ov} is about $2.27\times$, while Watt's algorithm~\cite{watt2023} requires at least $5$ full multiplications. Column $11$ shows that \cgbn{}~is faster than \text{CudaP} by large factors ($3.64$-$7.36\times$) for the lower precisions, but does not support precisions larger than $2^{15}$.

\text{F-Reg} exhibits smaller {\sc div-ov} than \text{CudaP} for all precisions other than $2^{16}$ and $2^{13}$. Moreover, column $7$ reports that for all precisions other than $2^{16}$, \text{F-Reg} is slower than \text{CudaP} by factors between $0.98$-$1.37\times$. Adjusting for \text{CudaP}'s advantage of efficient $128$-bit multiplication, it seems at least probable that with that, \text{F-Reg} performance would fall at least within $95\%$ of \text{CudaP}, if not overtake~it. 

The benefits of the register-placement optimization are evident in column $8$ that reports that \text{F-Shm} cannot run the highest four precisions, and is slower than \text{F-Reg} on the others by $1.31$-$2.37\times$ factors. As with multiplication, \text{F-Reg} can run precision $2^{19}$ with a scalable runtime, which is under $2\times$ that of precision $2^{18}$. 

\section{Conclusions and Future Work}
\label{sec:conclusions}
\spacetune{\balance}
We have provided high-level functional implementations of big integer addition, (classical) multiplication and (Newton iteration) quotient in Futhark.
These implementations use higher-order functions and resemble the mathematical formulations of the algorithms. 
The Futhark compiler generates code for GPUs, automating maps, function fusion and memory use to obtain efficiency.   

We have compared the execution efficiency of the resulting code to our hand-coded \cpp{} \cuda{} implementations of the same algorithms. The high-level Futhark versions for multiplication and quotient typically take about
$1.35\times$ as long as the hand-coded \cuda{} versions.   The Futhark version of addition runs at about the same speed or faster than the \cuda{} version.

We have also compared our implementations to the standard \cgbn{} library for big integer arithmetic on GPUs. 
While the Futhark versions are slower for small values, for larger values, the Futhark versions are faster for addition and multiplication ($>2^{14}$ bits for addition, $>2^{17}$ bits for multiplication). 
The performance gain increases as values get larger, for example at $2^{18}$ bits, addition was 
1.5 
to 3.5 
times as fast, 
multiplication was 26 
times as fast,
and \cgbn{} could not perform quotients.

This work has provided concrete examples leading to Futhark compiler improvements of general utility. Allocating arrays in register memory, when possible, has proven particularly useful.
We expect that further work on big integers will reveal other useful improvements.  A next step would be to implement more sophisticated multiplication algorithms.
\balance
\bibliographystyle{plain} 
\bibliography{main.bib}
\end{document}